\documentclass[12pt]{article}
\usepackage{epsf}
\begin{document}
\thispagestyle{empty}
\setcounter{page}{0}
\renewcommand{\theequation}{\thesection.\arabic{equation}}

{\hfill{\tt hep-th/0307024}}

{\hfill{KCL-MTH-03-08}}

{\hfill{ULB-TH/03-23}}

\vspace{1.5cm}

\begin{center} {\bf INTERSECTION RULES, DYNAMICS AND SYMMETRIES} 

\vspace{.5cm}

Fran\c cois Englert${}^a$, Laurent Houart${}^b$\footnote{Research Associate F.N.R.S.}
and Peter West${}^c$

\footnotesize \vspace{.5 cm}

${}^a${\em Service de Physique Th\'eorique\\ Universit\'e Libre de Bruxelles,
 Campus Plaine, C.P.225\\Boulevard du Triomphe, B-1050 Bruxelles, Belgium}\\ {\tt
fenglert@ulb.ac.be}

\vspace{.2cm}

${}^b${\em Service de Physique Th\'eorique et Math\'ematique }\\  {\em Universit\'e Libre de
Bruxelles, Campus Plaine C.P. 231}\\ {\em Boulevard du Triomphe, B-1050 Bruxelles,
Belgium}\\ {\tt lhouart@ulb.ac.be}

\vspace{.2cm}
 ${}^c${\em Department of Mathematics}, {\em King's College London}\\ {\em Strand, WC2R
2LS London U.K.}\\{\tt pwest@mth.kcl.ac.uk}

\end{center}

\vspace {1cm}
\centerline{ABSTRACT}
\vspace{- 3mm}
\begin{quote}\small We consider theories containing gravity, at most  one dilaton and form
field strengths. We show that the existence of particular BPS solutions of intersecting
extremal closed branes select the theories, which upon dimensional reduction to three
dimensions possess a simple simply laced Lie group symmetry $\cal G$.  Furthermore
these theories can  be fully reconstructed from the dynamics of such branes  and of their
openings. Amongst such theories are the effective actions of the bosonic sector of
M-theory and of the bosonic string. The BPS intersecting brane solutions form
representations of a subgroup of the group of Weyl reflections and outer automorphisms
of the  triple Kac-Moody extension
${\cal G}^{+++}$ of  the $\cal G$ algebra, which cannot be embedded in the overextended
Kac-Moody subalgebra ${\cal G}^{++}$ characterising the cosmological Kasner solutions. 
\end{quote}

\baselineskip18pt

\newpage

\setcounter{equation}{0}
\addtocounter{footnote}{-1}
\section{Introduction and conclusion}

 There is a widespread belief that a  consistent quantum theory of gravity and   matter
should emerge from some  non perturbative  M-theory generalisation of  superstrings  
admitting eleven dimensional supergravity as its low energy effective action. Upon
compactification,   its bosonic sector displays remarkable  symmetries which are often
viewed as the signature of   an underlying supersymmetry and are assumed to be
symmetries of the full would-be non perturbative   string theory. Supersymmetry controls
the existence of zero binding energy  bound states and leads to one of the most celebrated
result of the string theory approach to quantum gravity,   namely the correct counting of
states of some extremal and nearly extremal black holes built from extremal intersecting 
branes.

In this paper, we show that the existence of such zero binding energy    states in classical
general relativity is not a privilege of M-theory effective   actions. We find a large class of
theories, all of which contain gravity,  sharing these features.   Interestingly, these theories
include the low energy effective action of the bosonic string.

We consider generic theories in $D$ dimensions including gravity, one    dilaton and form field
strengths  of arbitrary degree and arbitrary couplings to the dilaton.  These   theories
admit solutions describing parallel extremal branes, i.e.  charged solitonic extended  
objects with  no force between them. Under precise circumstances there exists 
configurations   of closed extremal branes intersecting orthogonally in a configuration with
zero binding   energy. In this paper we will use the  BPS terminology for these
configurations\footnote{ BPS is used   here in  the original sense
 without referring to any supersymmetry property.}.  We will    review  the general
intersection rules \cite{aeh} which determine for which theories such  BPS   configurations
exist. This will lead to  constraints on  the dimension $D$, the admissible   field strengths
and their dilaton couplings. In particular cases the existence of such  BPS configurations
will allow a complete determination of the $D$-dimensional theory, including its
Chern-Simons terms, and will imply  that the
 dimensionally reduced theory possesses a Lie algebra symmetry.  

This happens in the following way.  We first show that the existence of   BPS intersecting
solutions with {\em one} common  direction between electric and magnetic 
$p$-branes ($p \geq 1$) pairs, stemming from the form field strength present in the
theory, is a   necessary condition to enhance the symmetry of the dimensionally reduced
theory to three space-time dimensions, from the torus deformation group $GL(D-3)$ to a
simply laced Lie group  
$\cal G$ of the same rank. We then consider    the possible appearance of open branes
terminating on closed ones. This possibility arises when   the intersection rules giving
  BPS configurations are such that  the number of dimensions on which   two extremal branes
intersect  is equal to the  boundary dimension of one of them. It is   known that in some 
theories, e.g. the theories describing the different phases of   M-theory, type IIA \cite{2a},
type IIB \cite{2b} and 11 supergravity  \cite{11}, consistency of  openings is ensured by the
presence of   Chern-Simons terms in the theory \cite{str,tow,aehw}. Here we will reverse
the logic. We   start with a theory which admits certain branes and by 
 analysing   the BPS configurations and demanding  that all possible openings  be realised
we find that the theory must possess  Chern-Simons terms. These terms may in turn   imply
the introduction of new form field strengths . This constructive process   can be iterated
until all possible openings are consistent with the Chern-Simons terms they   require.

In this construction the presence of BPS configurations will imply the existence of a simply
laced symmetry group   ${\cal G}$  in the theory  dimensionally reduced  to three 
dimensions. It is then inevitable  that  the theories we construct  in this way contain their
oxidation endpoints
\cite{cjlp}.  These are the  theories defined at the highest available space-time dimension
$D$ which upon   dimensional reduction to three  dimensions  are expressible as non linear
realisations by scalar   fields in a coset space
${\cal G}/{\cal H}$ where ${\cal H}$ is the maximal compact subgroup of  
${\cal G}$.

We have thus uncovered a  relation between brane physics and the presence   of symmetry. All
these maximally oxidised theories do possess the same type of BPS    configurations.  While
some of these theories, such as the bosonic sector of  eleven dimensional
supergravity, 
 can be extended to possess supersymmetry which 
  protects the BPS conditions against quantum corrections,  a number of  other theories do
not. However, all the theories    share very similar symmetries which may have important
consequences at the quantum level.   It has  been proposed that there is a huge algebraic
structure characterising   eleven-dimensional supergravity which is the very extended 
Kac-Moody algebra of the $E_8$   algebra, called $E_{11}$ (or
$E_8^{+++}$) \cite{west01}. The same very extended algebra occurs in IIA
\cite{west01} and IIB supergravity \cite{schnakw01}. Furthermore,  there has indeed been
evidence that such  a structure is realised, not only for supergravity, but     for its
proposed counterpart in the effective action of the bosonic string, namely
$k_{27}$ or  
$D_{24}^{+++}$ \cite{west01} and also for pure $D$ dimensional gravity where the proposed 
underlying symmetry algebra is thought to be 
$A_{D-3}^{+++}$ \cite{lw}. The notion of very extended algebras was introduced, and some of
their properties were discussed in reference \cite{ogw}.
It has been suggested that  all the maximally
oxidised theories,   possess   the very extension
${\cal G}^{+++}$ of the simple Lie algebra ${\cal G}$ \cite{ehtw}. The relation between branes
and symmetries found in this paper lends further support for these ideas. We indeed show  
that  the  BPS solutions   of the oxidised theory of a simply laced group ${\cal G}$ form
representations of a subgroup of the Weyl transformations of the algebra ${\cal G}^{+++}$.

The common BPS dynamical origin of such  hidden symmetries   in 
 maximally oxidised theories suggests that very extended Kac-Moody algebras generalise
the role played by supersymmetry in some of these theories.
 This is in line with the suggestion that fermionic strings and supersymmetry   originate in
the fermionic subspaces of the compactified bosonic string to   10-dimensional space-time
\cite{casherent85}. We recall that these subspaces accommodate all   D-branes of the
10-dimensional fermionic string theories \cite{englertht01}, yet they  provide no hint  
about the non-perturbative origin of the
$NS5$-brane out of the bosonic string. The BPS configuration of the electric string and its
21-magnetic dual  in the bosonic string effective   action considered here in Section 2
allows to view these objects, upon such compactification, as  parents of the fermionic BPS  
string and its
$NS5$-dual.

The paper is organised as follows. In Sections 2 and 3, we review and illustrate
respectively the intersection rules for extremal closed branes and the conditions allowing
their openings. In Section 4, we relate the possible emergence of symmetry in  dimensional
reduction to the intersection rules. We show that the existence of BPS configurations 
consisting of extremal electric $p$-branes and of their magnetic duals intersecting along
one common dimension constitute  a necessary condition for having a maximally  oxidised
theory of a simple simply laced Lie group $\cal G$. In Section 5, we use the opening of
branes to generate Chern-Simons terms and effectively reconstruct {\em all} such
theories. In Section 6, we relate the existence of the BPS solutions to the group of of Weyl
reflections and outer automorphisms of the triple Kac-Moody extensions  ${\cal G}^{+++}$ of
the algebra of $\cal G$.

\setcounter{equation}{0}
\section{Intersection rules}

In this section we will review along the line of ref.\cite{aeh}   the general rule determining
how extremal branes can intersect orthogonally  in a configuration with zero binding
energy.

We begin with a generic theory in 
$D$ dimensions  which includes gravity, a dilaton and ${\cal M}$ field strengths  of arbitrary 
form degree $n_I$  with $n_I\leq D/2$  and arbitrary couplings to the dilaton $a_I$. If  $n_I > 
D/2$  we replace the field strength in the action by its dual defined by
\begin{equation}
\label{hodge}
\sqrt{-g} e^{a\phi}F^{\mu_1 \dots \mu_n}={1\over (D-n)!}
\epsilon^{\mu_1 \dots \mu_n \nu_1 \dots \nu_{D-n}} 
\tilde{F}_{\nu_1 \dots \nu_{D-n}}\, .
\end{equation}
The action is
\begin{equation} S= \frac{1}{16\pi G_N^{(D)}} \int d^D x \sqrt{-g}
\left( R - \frac{1}{2} (\partial \phi)^2 -
\sum_I \frac{1}{2 n_I !}e^{a_I \phi}F_{n_I}^2 \right)\, , \quad I=1\dots {\cal M}\, , \label{action}
\end{equation} where we have not included  possible  Chern-Simons terms.   It  can be shown
that they have no effect on  the zero binding energy solutions considered here although such
terms will be important in the following sections. This omission apart, the above action
describes the  bosonic sectors of all supergravity theories, the prototype examples being
eleven dimensional supergravity
\cite{11},  and the IIA
\cite{2a} and  IIB \cite{2b} supergravities.  In this paper,  we will also consider  theories 
which do not possess supersymmetry but which are still described by Eq.(\ref{action}). 

The zero binding energy configurations of closed extremal branes intersecting
orthognally  are obtained by first specialising to metrics of the following diagonal form,
\begin{equation} ds^2=-B^2 dt^2+\sum_{i=1}^pC_{(i)}^2(dy^i)^2+ \sum_{a=2}^{D-p} G^2
(dx^a)^2
\, ,
 \label{metric}
\end{equation} where $y^i$ are compact coordinates and the functions $B$, $C_{(i)}$,  $G$
depend only on the overall transverse coordinates $x^a$ and we allow for multicenter
solutions. We choose $p$
so that all branes are wrapped in the compact dimensions and that no compact dimension is
transverse to all branes. The overall transverse space is non-compact and has dimension 
$D-p-1$.  We  recall that for each brane of dimension $q$ present in the solution with $q$
less than
$p$,  we have to take, in the  compact directions transverse to
the brane, a ``lattice'' of such $q$-branes  and then to average over them.

For the $n$-form field strengths we may choose two different ans\"atze,
\begin{eqnarray}
\mbox{Electric}& :& F_{ti_1\dots i_{q_A}a}=\epsilon_{i_1\dots i_{q_A}}\partial_a E_A\, ,
\label{electric}\\
\mbox{Magnetic}& :& \tilde{F}_{ti_1\dots i_{q_A}a}=\epsilon_{i_1\dots i_{q_A}}\partial_a
\tilde E_A\, .
\label{magnetic}
\end{eqnarray} The space-time charges are thus respectively defined by
\begin{equation} Q^{el}_A\propto \int *F_{q_A+2}\, , \qquad Q^{mag}_A\propto \int
F_{D-q_A-2}\, .
\label{charges}
\end{equation} Here, the dual $*F$ is defined by Eq.(\ref{hodge}) and $A=1\dots{\cal N}$,
where
${\cal N}$ is the total number of electric and magnetic distinct non-parallel branes. This
number can of course exceed the number of different $n$-forms.

We  recall briefly how in reference \cite{aeh}  solutions of the equations of motion of the
general action Eq.(\ref{action})  describing zero binding energy configurations between
${\cal N}$ distinct non-parallel extremal branes were found.      Two  ans\"atze where made
in order to ensure the   no-force condition between the constituent branes. The first  one,
which considerably simplifies the equations of motion, is 
\begin{equation}
 BC_1\dots C_p G^{D-p-3}=1\, .
\label{ansatz}
\end{equation}  The second expresses that one independent harmonic function  is
associated to  each brane, and that the solution is completely characterised by  ${\cal N}$
harmonic functions
$H_A$ with  $E_A \propto H_A^{-1}$ and $A=1 \dots {\cal N}$, where we have dropped the tilde
distinguishing electric and magnetic  ans\"atze.

The  Einstein equations for the diagonal ${R^a}_a$  components and  the dilaton equations
yield the following expressions for an overall transverse space\footnote{For $D-p \le 3$,
space is not asymptotically flat but the solution is still characterised by harmonic
functions and the intersection rule equations Eq.(\ref{intrule}) below still apply.}
$D-p >3$,
\begin{equation} ds^2=-\prod_A H_A^{-2\frac{D-q_A-3}{\Delta_A}}dt^2+\sum_{i=1}^p
\prod_A H_A^{-2\frac{\delta^{(i)}_A}{\Delta_A}}(dy^i)^2+ \sum_{a=2}^{D-p}
\prod_A H_A^{2\frac{q_A+1}{\Delta_A}}(dx^a)^2\, , \label{metric_sol} 
\end{equation}
\begin{equation}  e^{\phi}=\prod_A H_A^{\varepsilon_A a_A\frac{D-2}{\Delta_A}}\, ,
\label{dilaton_sol} \end{equation} with 
\begin{equation} H_A=1+\sum_k \frac{c_A Q_{A,k}}{|x^a-x^a_k|^{D-p-3}}\, .
\label{multicenter}
\end{equation} where the $Q_{A,k}$ are the charges of the branes located at $x_k^a$ and the
$c_A$ are constants. 
$\Delta_A=(q_A+1)(D-q_A-3)+\frac{1}{2}a_A^2(D-2)$ , 
$\varepsilon_A=+(-)$ for electrically (magnetically) charged branes, and
$\delta^{(i)}_A=D-q_A-3$ or $-(q_A+1)$ depending on whether $y_i$ is parallel or
perpendicular to the
$q_A$-brane. In order to build up the metric, we  thus  include, for each  $q_A$-brane in the
configuration, a  factor  
$H_A^{-2\frac{D-q_A-3}{\Delta_A}}$ in front of each coordinate (including the time
coordinate) longitudinal to the brane, and a factor of 
$H_A^{2\frac{q_A+1}{\Delta_A}}$ in front of each transverse coordinate.  This is in 
agreement with the harmonic superposition rule  formulated first in reference \cite{tsey} in the
particular context of M-theory. The solution above Eq.(\ref{metric_sol}) is asymptotically
flat and  the total mass of such a configuration is,  as expected, the sum of the masses of
each constituent brane, which are equal to the charges:
$M=\sum M_A = \sum Q_A$.

In this model independent way of deriving zero binding energy  configurations we get an
important bonus namely  the  {\it intersection rule equations} \cite{aeh}.  We  did not yet
use the
${R^a}_b$ off-diagonal  components of the Einstein equations and  these reduce to a set of
algebraic  conditions. These  are the intersection rules equations which yield, for each  pair
$(A,B)$ of distinct 
$q$-branes of dimensions $(q_A, q_B)$,       the number of dimensions $\bar q$ on which they
intersect in terms of the total number of space-time dimensions $D$ and of the field
strength couplings to the dilaton. They read
\begin{equation}
\bar{q}+1=\frac{(q_A+1)(q_B+1)}{D-2}-\frac{1}{2}
\varepsilon_A a_A \varepsilon_B a_B\, . \label{intrule}
\end{equation}  Thus, in every theory whose action has the form shown in Eq.(\ref{action}), a 
zero  binding energy configuration
$(M=\sum M_A = \sum Q_A)$ between any chosen set of extremal branes of the theory 
exists provided the pairwise conditions Eq.(\ref{intrule}) yield   integer\footnote{ The case
$\bar{q}=-1$ is  relevant and can be interpreted in  terms of instantons in the Euclidean. In
that case the time coordinate   need not be longitudinal to all  branes.}  $\bar{q}$ in the
range
$(-1\leq {\bar q} \leq q_A,q_B$).  The configuration is then described by the metric 
Eq.(\ref{metric_sol}), the dilaton Eq.(\ref{dilaton_sol}), and the field strength expressed in
terms of the $H_A$ given in Eq.(\ref{multicenter}).

It is interesting to point out some properties of the pairwise intersection rule equations.  
Consider a theory containing a form $F_{n_A}$ and hence  extremal electric 
$q^e_A$-brane and  magnetic $q^m_A$-brane  $(q_A^m=D-q_A^e-4)$  solutions and a form
$F_{n_B}$ and hence corresponding $q^e_B$ and $q^m_B$-branes.  It  is easy to see from
Eq.(\ref{intrule}) that if one of  the four possible  intersections between a $q_A$-brane  and
a
$q_B$-brane has integer  ${\bar q}$,  then  the three other  intersections have also integer
dimensions. We have indeed
\begin{eqnarray} {\bar q}^{(e_A,e_B)}&=&q^e_A-{\bar q}^{(e_A,m_B)}-1\, , \nonumber\\
{\bar q}^{(m_A,m_B)}&=&D-q^e_B-{\bar q}^{(e_A,m_B)}-5\, ,\label{qbar}\\ {\bar
q}^{(m_A,e_B)}&=&q^e_B-q^e_A+{\bar q}^{(e_A,m_B)}\, , \nonumber
\end{eqnarray} where in ${\bar q}$ the first superscript labels  the electric or  magnetic
nature of the $q_A$-brane and  the second that of the $q_B$-brane. In particular, putting
$A=B$, the relations Eq.(\ref{qbar}) are valid for the branes arising from a single form. 

The intersection rules reviewed here were applied to the different ``phases'' of M-theory
(11D supergravity, type IIA and type IIB theories)  in \cite{aeh}. They gave back the
well-known zero binding energy  configurations  preserving  some supersymmetries. These
brane configurations  were originally derived from  supersymmetry and duality arguments
(see for instance \cite{allir} and reference therein)\footnote{  In \cite{tsey2} a derivation
of the intersection rules not based on  supersymmetry has been given for the branes of
M-theory  using brane probes in brane backgrounds.}.  The generality of the intersection
rules allows `BPS'
 configurations of extremal branes with vanishing binding energy in the broader context 
where
 supersymmetry is not required. 

We now illustrate by one important example the emergence of such  BPS states in
non-supersymmetric theories. We take the effective action Eq.(\ref{action}) of  the bosonic
string (omitting  the tachyon field) generalised to $D$ dimensions.  The corresponding
string theory is consistent at the quantum level only for $D=26$ but in the present context
nothing prevent us to extend the
  action  Eq.(\ref{action}) to any dimension $D$, keeping   only a 3-form\footnote {Since we
have chosen $n_I
\leq D/2$  we have here $D \geq 6$.} denoted  in what follows $H_3$. One has
\begin{equation} S= \int d^D x \sqrt{-g}
\left( R - \frac{1}{2} (\partial \phi)^2 -
\frac{1}{12}e^{a_3 \phi}H_3^2 \right)\, , \label{dnaction}
\end{equation} with the dilaton coupling $a_3$ given by
\begin{equation} a_3= \sqrt{8\over D-2}\, .
\label{dndilaton}
\end{equation}  The 3-form $H_3$ gives rise to an  extremal electric 1-brane
 describing the fundamental string,  traditionally denoted  $1F$, and to its  dual $D-5$
magnetic brane that we denote $NS(D-5)$. We  apply the intersection rules to seek for
possible BPS  states of intersecting extremal branes in the  model.  Using
Eq.(\ref{dndilaton}) in Eq.(\ref{intrule})  we immediately see that a BPS solution exists for
the three  possible pairwise intersecting configurations.  We have 
\begin{equation}  1F\cap 1F=-1 \qquad 1F\cap NS(D-5) = 1 \qquad NS(D-5)\cap
NS(D-5) =D-7\, , 
\label{dnbranes}
\end{equation} where we used the notation  $q_A \cap q_B =\bar{q}$. The corresponding BPS
solutions are given by Eq.(\ref {metric_sol}) and Eq.(\ref{dilaton_sol}) with dilaton coupling
Eq.(\ref{dndilaton}). From the pairwise rules we can then construct BPS configurations with
more than two branes.  Applying these results to  the  bosonic string theory,  namely
taking $D=26$, we can
 build BPS configurations between various $1F$ and $NS21$.

\setcounter{equation}{0}
\section{The opening of branes}

In this section we   analyse the  breaking of closed extremal branes into open branes
terminating on closed ones. We  consider the BPS configurations given by Eq.(\ref{intrule})
in the special case when
${\bar q}$ has the  same dimension as the potential boundary of one of the two constituent
branes, i.e
$q_A-1$ or
$q_B-1$, and study its possible opening.  Such openings require  the addition of
Chern-Simons terms to the action Eq.(\ref{action}) and may enlarge the brane content of the
theory. We shall see in Section 5 that, under some conditions, such openings  fully
determine   the theory and  relate brane dynamics to the existence of a symmetry. The
presentation given in this section is a generalisation of the one performed in the context
of M-theory \cite{str,tow,aehw}.

Let us review  how extended objects  carrying a conserved charge can be opened. The main
obstacle towards  opening of branes is charge conservation. Generically,  the charge
density of a
$q$-brane is measured by performing an integral of the relevant field strength on a
$(D-q-2)$-dimensional sphere $S^{D-q-2}$ surrounding the brane in its transverse space,
\begin{equation}  Q_q \propto \int_{S^{D-q-2}} *F_{q+2}\, .
\label{charge}
\end{equation}
 If the brane is open, we can slide the $S^{D-q-2}$ off the loose end and shrink it to zero
size. This would imply the  vanishing of the charge and hence a violation of charge
conservation. This conclusion is avoided if, in the above process, the
$S^{D-q-2}$  necessarily goes through a region in which the equation
\begin{equation}  d*F_{q+2}=0
\label{fmot}
\end{equation} no longer holds.  We shall see that this is  the case when the open brane ends
on some other  one.

In the framework of M-theory  the source terms needed in Eq.(\ref{fmot}) 
to ensure charge conservation for the open branes originate from two requirements
whose interplay  leads to a consistent picture. 
On the one hand there are space-time
Chern-Simons type terms in supergravity which allow for charge conservation for well
defined pairing of open and `host' branes
\cite{tow}. On the other hand   the world-volume effective actions
\cite{str} for the branes of M-theory relate world-volume fields and  pull-backs of
space-time fields, and  gauge invariance \cite{wit} for open
branes ending on the `host' brane implies that the end of the open branes acts
as a source for the world-volume field living on the closed `host' brane. 

In reference \cite{aehw} a systematic study in M-theory of all the zero binding energy
configurations Eq.(\ref{intrule}) corresponding to 
${\bar q} = q_A-1$ (with $q_A \leq q_B$) was performed. It was shown that in all cases it
was possible to open the $q_A$-brane along its intersection with the $q_B$-brane. The
crucial ingredient was the presence of the  appropriate Chern-Simons terms in the
supergravity Lagrangians for each case.

Here we propose to reverse the logic. Starting with a Lagrangian of type Eq.(\ref{action})
and having zero binding energy configurations between branes we will ask that, if
${\bar q}=q_A-1$,   the corresponding $q_A$-brane  open  on the
$q_B$-brane.  This will {\it determine} the form of the Chern-Simons terms one has to add
to Eq.(\ref{action}) and will also  in some cases require the {\it introduction of new
field strength  forms} 
$F_{n_I}$. One then proceeds iteratively. 

We illustrate the role of the Chern-Simons term (see also \cite{tow,
aehw})  by taking as example    a theory  with only one $n$-form  $F_{q^e+2}$ and
dilaton coupling in Eq.(\ref{action})  such that  the intersection rule between the
electric $q^e$-brane and the  magnetic
$q^m$-brane ($q^m=D-q^e-4$) is ${\bar q}=q^e-1$.

We  modify the Eq.(\ref{fmot}) for $F_{q^e+2}$ in order to be able to allow the opening of $q^e$
on $q^m$ by the addition   to the action Eq.(\ref{action}) of the Chern-Simons term
\begin{equation}\label{cs}
\int A_{q^e+1}\wedge F_{q^e+2}\wedge F_{D-2q^e-3}\, ,
\end{equation}
and of a  standard   kinetic energy term for the  new\footnote{ If $D=3q^e+5$
there is no need to introduce a new field strength as the Chern-Simons terms can be build with
the field strength
$F_{q^e+2}$.}  field strength  $F_{D-2q^e-3}$. Its dilaton coupling is chosen
such that the intersection rules Eq.(\ref{intrule})
give  integer  intersection dimension between the new extremal electric
$(D-2q^e-5)$-brane and its dual magnetic
$(2q^e+1)$-brane.   Charge conservation now reads
\begin{equation}  d*F_{q^e+2} = F_{q^e+2} \wedge F_{D-2q^e-3} + Q^e
\delta_{D-q^e-1}\, . \label{emcase}
\end{equation} Here and in what follows wedge products are defined up to signs and 
numerical factors. In the r.h.s. of Eq.(\ref{emcase}) the first term comes from the  variation
of the   Chern-Simons term and  the second one is the $q^e$-brane
charge density. 
$\delta_{D-q^e-1}$ is the Dirac delta function in the directions transverse to the
$q^e$-brane. We  introduce here an explicit source term for the electric brane since, to
study its opening, we want to extend to the
branes themselves the validity of the usual closed brane solution .  Such term is required
because the equations of motion from which the intersecting brane solutions were
derived do not contain any source term and are therefore valid only outside the sources.

Now it is easy to see that  opening is  consistent with charge conservation.
Taking into account that in the configuration considered  no brane is associated to the new
field, we have
\begin{eqnarray} F_{D-2q^e-3}&=&dA_{D-2q^e-4}\, ,\\
dF_{q^e+2}&=&Q^m
\delta_{q^e+3}\, ,
\end{eqnarray} and one can rearrange Eq.(\ref{emcase}) in the following way,
\begin{equation} d( * F_{q^e+2}- F_{q^e+2}\wedge A_{D-2q^e-4})=Q^e \delta_{D-q^e-1} 
		- Q^m \delta_{q^e+3}\wedge A_{D-2q^e-4}\, .
\label{emcase2}
\end{equation} 

A $S^{D-q^e-1}$ sphere, which has only one point in common with
 the open $q^e$-brane, intersects the $q^m$-brane on a
$S^{D-2q^e-4}$ sphere surrounding the intersection. Integrating Eq.(\ref{emcase2}) over the
$S^{D-q^e-1}$ sphere, one gets
\begin{equation}
0=Q^e- Q^m \int_{S^{D-2q^e-4}}A_{D-2q^e-4}\, .
\label{chcons} 
\end{equation}
This equation can be rewritten as:
\begin{equation} Q^e=Q^m . Q_1
\qquad \mbox{with} \qquad  Q_1\equiv\int_{S^{D-2q^e-4}}A_{D-2q^e-4}\, .
\label{effcharge} 
\end{equation} We see that, up to a closed form $da_{D-2q^e-3}$ on the closed
$q^m$-brane, the pull-back
$\hat A^{(q^m)}_{D-2q^e-4}$ of the potential
$A_{D-2q^e-4}$ to this brane  behaves like a
$D-2q^e-4$-form field strength, magnetically coupled to the boundary.  We thus write the
field strength on the $q^m$-brane as
\begin{equation}\label{calGdef} {\cal G}_{D-2q^e-4}=\hat
A^{(q^m)}_{D-2q^e-4}-da_{D-2q^e-3}\, .
\label{gfield}
\end{equation}
 Equivalently, one may formulate  the world-volume theory of the closed
`host' brane  in terms of the world-volume Hodge dual ${\cal
F}_{q^e+1}\stackrel{def}{=}{}^\star {\cal G}_{D-2q^e-4} $, thus interpreting  the charge
$Q_1$ at the end of the $q^e$-brane as an electric charge on the $q^m$-brane.

The above argument applied to the electric brane source of the new field
$F_{D-2q^e-3}$ shows that the new extremal electric brane can also be opened on the host
closed magnetic brane $q^m$, provided the intersection rules are satisfied between the
new and the original branes, as will indeed be the case for the problem studied in Section 5.
 Its end carries a magnetic charge $Q_2$ which is the source  of a field ${\cal
H}_{q^e+1}$ on the brane which may tentatively be identified\footnote{Strictly speaking,
this identification is not necessary for what follows but appears natural as it avoids a
doubling of fields degrees of freedom on the brane
$q^m$. In the M-theory case, this identification can be proven \cite{aehw} and we surmise
that it is a general consequence of the symmetries uncovered in this paper. Note that this
identification implies that the field  $f=da$ in Eq.(\ref{calGdef}) {\em must} be present  on
the world volume of the
$q^m$-brane  to ensure  gauge invariance \cite{wit}  for the open  branes.} with
${\cal F}_{q^e+1}\stackrel{def}{=}{}^\star {\cal G}_{D-2q^e-4} $.

The world-volume point of view  gives a  picturesque way to determine  which branes have
to be added in order to have a consistent opening. Indeed,  when a brane opens on a closed
`host' brane,  the boundary appears from the world-volume point of view  as a charged
object  under a world-volume field strength living  in the closed brane. The world-volume
Hodge dual of this object is   the boundary of an other brane which can also be
consistently opened on  the same closed  `host'  brane. The field strength associated to
this new brane is precisely the one appearing in the Chern-Simons ensuring the consistency
of the opening of the  brane we started with. 

In summary, having  a theory  of type Eq.(\ref{action})  in which some zero binding energy
configurations  give potential boundaries, it is possible to complete it in a well-defined
way (adding Chern-Simons terms\footnote{The coefficients of the Chern-Simons terms
Eq.(\ref{cs}) are not fixed in this qualitative discussion. Their precise values are  important
when one will discuss the potential symmetries.  In the framework of M-theory they are usually
fixed by supersymmetry.  Nevertheless it is possible to fix them or at least quantise them
generically using only consistency arguments without appealing to supersymmetry, see for
instance \cite{bachas, bg}.},  and when necessary new form field strengths, hence
new branes) in order to ensure consistency of brane  opening with charge conservation. In
section 5, such dynamical requirement will be at work to  reconstruct theories characterised
by a coset symmetry, when dimensionally reduced,  purely from brane considerations.

\setcounter{equation}{0}

\section{D=3 cosets and intersection rules}

It has been realised long ago that the scalars  occurring in supergravity theories belong to
cosets or non-linear realisations of a Lie group.  This is a consequence of supersymmetry, but
the Lie algebras that describe the cosets were not anticipated. They have been the subject of
much study, and some classic examples are given in 
\cite{allcoset}. The more scalars one has, the bigger the corresponding Lie algebra.
Consequently,  when one  performs a dimensional reduction of a given supergravity theory
on tori, 
 the algebras of the non-linear realisation  grow in size corresponding to growth in the
number of  scalars  in the dimensional reduction process.  The full supergravity theories
obtained in this way always admit  a 
 non-compact Lie group symmetry $\cal G$ which     
  is manifest in the scalar sector of the theory.  This sector is often described by 
 a non-linear realisation of the group in the form of a coset space ${\cal G}/{\cal H}$   where
$\cal H$ is the maximal compact subgroup of ${\cal G}$. The finite dimensional Lie group
$\cal G$ 
 reaches its largest rank  when  the  dimensional reduction is performed all the way down to
three  dimensions, since  all the original fields can then indeed be expressed in terms  of
scalars. The complete  Lagrangian is  then a conventional non-linear realisation of scalar
fields.

  For each  finite dimensional semi-simple Lie group
$\cal G$ one can consider the corresponding three dimensional  scalar coset and  in
reference
\cite{cjlp}  a higher dimensional theory  which leads  upon dimensional reduction to the
three dimensional scalar coset theory is found for each
$\cal G$. This process is called oxidation and the theory in the highest dimension which
leads to the three dimensional  scalar coset for a given
$\cal G$ is referred to as a  maximally oxidised theory.  In fact, the dimensional reduction of
a generic field theory leads 
  to a theory  with no Lie algebra symmetry.  However, there are   examples
of theories that possess no supersymmetry,  but whose dimensional reduction leads to
scalars that belong to non-linear realisations. Perhaps the best known is the dimensional
reduction of   pure gravity in any dimension. The occurrence and the uniqueness of a
symmetry resulting from dimensional reduction were investigated in reference  \cite{lw}
and one can examine in detail how the resulting Lie group symmetry  in three
space-time dimensions places relations  between the field content and couplings
of the higher dimensional theory. 

In this section we will perform the dimensional reduction of  a  generic theory described by
actions of the type  Eq.(\ref{action}),  with  the option to add some Chern-Simons terms
when necessary.  We will discuss the possible emergence of a coset symmetry  in three
dimensions. Some particular cases presented here have already been considered   in
\cite{lw} whose line of argument we follow.   We will relate the possible emergence of 
symmetry to the intersection  rules and thus  to the possible existence of BPS
configurations between  extremal branes characterising a theory. We will uncover a
precise  relation between the intersection rules and the onset of a symmetry.

To fix  notations in a self-contained presentation, we start   by recalling the well-known
dimensional reduction method. We will follow closely the  discussion  outlined in references
\cite{pope} and \cite{lw}. 

Starting with a theory defined in $D$ dimensions  by an action  of the type
Eq.(\ref{action}), we  compactify  to $D-1$  dimensions and,  remaining  in the
Einstein frame with  the standard 1/2  kinetic term normalisation for the new scalar.   This
procedure amounts to take as compactification ansatz\footnote{This corrects a typo in
Eq.(2.1) of  ref.\cite{lw}.}
\begin{equation} ds^2_D = e^{2\beta_{D-1}\phi_2}ds^2_{D-1}  +
e^{-2(D-3)\beta_{D-1}\phi_2}(dx^{D-1}+A_\mu dx^\mu)^2\, .
\label{mcomp}
\end{equation} We used the notation $\phi_2$ for the scalar appearing in the first step of
the dimensional reduction and we rename 
$\phi_1$   the dilaton $\phi$   already present in the uncompactified theory
defined by Eq.(\ref{action}). The compactified coordinate is $x^{D-1}$, $\mu=0 \dots D-2$, and
\begin{equation}
\beta_D = \sqrt{1\over2(D-1)(D-2)}\, .
\label{alpha}
\end{equation} The gravitation part of the action Eq.(\ref{action}) becomes
\begin{eqnarray}
\int d^D x \sqrt{- g_D} \,  R_D &=& \int  d^{D-1} x  
\sqrt{-g_{D-1}}\, (\,  R_{D-1} - {1 \over 2}
\partial_\mu\phi_2\partial^\mu\phi_2
\nonumber \\ &-& {1\over 4}e^{-2(D-2) \beta_{D-1}\phi_2}\, F_{\mu\nu}F^{\mu\nu}\, )\, .
\label{gredu}
\end{eqnarray} where $ F_{\mu\nu}=\partial_\mu A_\nu - \partial_\nu A_\mu$.

For each $n_I$-form field strength
$F_{n_I}$ in Eq.(\ref{action}) we get after reduction
\begin{eqnarray}
\int d^{D} x\, \sqrt{-g_D}\, {1 \over 2 n_I!} \, e^{a_I  
\phi_1} F_{n_I}^2  &=& \int d^{D-1} x  \sqrt{-g_{D-1}}\, ( \, {1 \over 2 n_I!}\, e^{a_I 
\phi_1-2(n_I-1)\beta_{D-1}\phi_2}\,  F_{n_I}^{{\prime}2} \nonumber \\  &+&   {1\over 2
(n_I-1)!}\,  e^{a_I  \phi_1 + 2(D-1-n_I) \beta_{D-1} \phi_2}\,  F_{n_I-1}^2 \, )\, ,
\label{freduc}
\end{eqnarray} where 
\begin{equation}
 F_{\mu_1 \dots \mu_n}^{\prime}= F_{\mu_1 \dots \mu_n}-
 n\, F_{ [ \mu_1 \dots \mu_{n-1}} A_{\mu_n ] }\, .
\label{fmod}
\end{equation}

We can repeat this procedure step by step to obtain the theory on a $p$-torus. One has then
obviously $p$ scalars $\phi_j$ with 
$j=2 \dots p+1$  parametrising the radii of the torus,   coming from  the diagonal
components of the metric in the compact dimensions. Additional scalars denoted
$\chi_{\vec\alpha}$ arise  from several origins. They come from  potentials $A^k_\mu$  
which arise when reducing gravity from $D+1-k$ to $D-k$  (see Eqs.(\ref{mcomp}) and
(\ref{gredu}) ) and also from  the potentials associated to the $F_{n_I}$ (when $p \geq
n_I-1$) with indices in the compact dimensions. In addition the $n$-form field strengths give
additional scalars when
$p=D-n-1$ by dualising them. In particular when we reach $D=3$ the  $F^k_{\mu \nu}$  (with
$k=1
\dots D-3$) coming from the gravity part of the action (i.e. the graviphotons) can be
dualised to scalars, and we are left with only  scalars. The action takes then the 
form
\begin{equation} S = \int d^3 x \, \sqrt{-g_3} \, 
\left(R_3 - {1 \over 2} \partial_\mu \vec{\phi}\cdot\partial^\mu \vec \phi - {1\over2}
\sum_{\vec
\alpha}e^{\sqrt{2} \vec\alpha \cdot \vec\phi}
\partial_\mu\chi_{\vec\alpha}\partial^\mu\chi_{\vec\alpha}+\dots\right)\, ,
\label{clag}
\end{equation} where $\vec\phi = (\phi_1,...,\phi_{D-2})$, the $\vec\alpha$ are 
constant
($D-2$)-vectors\footnote{ The normalisation factor
$\sqrt{2}$ has been chosen for convenience. It will eventually correspond in the simply
laced case to the standard normalisation of the roots, namely 
$\vec\alpha \cdot \vec \alpha =2$.} characterising  each $\chi_{\vec\alpha}$. If we start in
$D$ dimensions without dilaton the vectors are of course ($D-3$)-dimensional as $\phi_1$ is
absent in that  case. The ellipsis in Eq.(\ref{clag}) stands for terms of order higher than
quadratic in the 
$\chi_{\vec\alpha}$ scalars. They come from the modification of the field strengths   
Eq.(\ref{fmod}) in the dimensional reduction process and also from 
  possible Chern-Simons terms   in the uncompactified theory.

We  now   recall  in which circumstances a Lagrangian  Eq.(\ref{clag})  can be identified with a
non-linear realisation of  ${\cal G}$ in a coset space
${\cal G}/{\cal H}$  \cite{pope}.  More precisely, we  consider here a
split (maximally non-compact) group 
${\cal G}$    with  generators ${\vec
H}$ and
$E^{\vec\alpha}$  where the Cartan subalgebra is generated by 
$\vec H$ and 
\begin{equation} [\vec H, E^{\vec \alpha}] = \vec \alpha E^{\vec \alpha}\, .
\label{cartanb}
\end{equation} Roots split  into positive and negative
ones. In what follows we call a root   positive (negative) if its first non-vanishing
component, as counted from the right, is positive (negative).  The positive roots can be
written as linear combinations, with non-negative integer coefficients, of the so-called
simple roots. The Cartan matrix,  which uniquely determines the algebra of ${\cal G}$, is
defined in terms of the simple roots and is given by
\begin{equation}
A_{ij}=2{\vec\alpha_i\cdot\vec\alpha_j\over\vec\alpha_i\cdot\vec\alpha_i}\, .
\label{cartanm}
\end{equation}
The  Cartan involution
$\tau:(E^{\vec\alpha},E^{-\vec\alpha},\vec H)
\rightarrow -(E^{-\vec\alpha},E^{\vec\alpha},\vec H)$ can  be used to construct the maximal
compact subgroup ${\cal H}$ as the subgroup    invariant under the involution.   We can then
construct
${\cal G}/{\cal H}$ non-linear realisations. We write the coset representatives  by
exponentiating the Borel subalgebra generated by the Cartan and the positive root
generators, namely,
\begin{equation} {\cal U} = e^{{1 \over \sqrt{2}}\vec \phi \cdot \vec H}
\sum_{\vec \alpha >0}e^{\chi_{\vec \alpha}\cdot E^{\vec\alpha}}\, ,
\label{Udef}
\end{equation} where the sum is only over the positive  roots. One can show that the
scalar Lagrangian
\begin{equation} {\cal L}_{\cal G/H} = {1\over 4}{\rm Tr}\left(
\partial_\mu{\cal M}^{-1}\partial^\mu{\cal M}\right)\, ,
\label{cosetl}
\end{equation} where
\begin{equation} {\cal M} = {\cal U}^\#{\cal U}\, ,
\label{Mdef}
\end{equation} is invariant under global ${\cal G}$ transformations and local ${\cal H}$
transformations ${\cal U}\rightarrow  h  {\cal U}  g$.  Here we  used the generalised
transpose acting on the generators: 
$X^\#=\tau(X^{-1})$. If we normalise the Cartan and positive-root generators so that 
\begin{equation} {\rm  Tr}(H_iH_j)=\delta_{ij}\qquad  {\rm  Tr}(E^{\vec\alpha}
E^{\vec\beta})=0\quad {\rm  Tr}\left( {E^{\vec\alpha}}
E^{-\vec\beta}\right)=\delta^{\vec\alpha\vec\beta}\, ,
\label{norm}
\end{equation}  one can show that ${\cal L}_{\cal G/H}$ is precisely   the
scalar  part of the Lagrangian in the action Eq.(\ref{clag}).

Thus it follows that the action Eq.(\ref{action}) when dimensionally reduced to three
dimensions  has a ${\cal G}/{\cal H}$  symmetry {\it if} the vectors $\vec\alpha$ obtained
from the compactification can be identified with the positive roots of a group ${\cal G}$ and
{\it if}, when necessary,  some precise Chern-Simons terms are added in the
uncompactified  theory. Adding some precise Chern-Simons terms in Eq.(\ref{action}) may
be indeed  required in order to match the terms obtained in the dimensional reduction,  which
are included in  the ellipsis  in Eq.(\ref{clag}),   with
${\cal L}_{\cal G/H}$ (see for instance
\cite{pope}). The requirement that the 
$\vec\alpha$ correspond to positive roots is thus a necessary condition to uncover a
symmetry.

We are now ready to discuss  the dimensional reduction of  a generic theory described by
the action Eq.(\ref{action}) and  the  necessary conditions  for the emergence of a coset
symmetry in  three dimensions. The discussion generalises the one of ref.\cite{lw}. We will then
uncover the relation between the onset of symmetry and the intersection rules.

We first recall  the well-known  dimensional reduction  of pure gravity
(see Eq.(\ref{gredu})) down to three dimensions, which  leads to ${\cal G}= SL(D-2)$ whose
algebra is
$A_{D-3}$.  The scalars corresponding to the simple roots of $A_{D-3}$ are of two kinds. 

There are first $D-4$ scalars which are the components $A^k_{D-k-1}$ of the potentials 
coming  from
$g_{D-k,D-k-1}, \ k=1 \dots D-4$.   These are obtained by performing the ``fastest''
reduction  on the potentials $A_\mu^k$ (see Eq.(\ref{gredu})) to obtain  a scalar going from
$D-k$ to $D-k-1$ when compactifying on $T^{k+1}$.  The corresponding simple roots 
$\vec{\alpha}^g_k$ are given by
\begin{eqnarray}
\vec\alpha^g_k &=& \sqrt{2}(0;
\underbrace{0, \dots , 0}_{k-1 \,   {\rm terms}} \, , - (D-k-1)\beta_{D-k}, 
(D-k-3)\beta_{D-k-1},
\underbrace{0, \dots ,0}_{D-4-k\, \rm{terms}})\, ,
\nonumber \\ k&=& 1 \dots D-4\, .
\label{gchain}
\end{eqnarray} They define a subalgebra  $A_{D-4}$. We have indeed
\begin{equation}
\vec\alpha^g_k\cdot\vec\alpha^g_l = \left\{\matrix{ 2\cr -1\cr 0\cr} \ 
\matrix{k=l\cr|k-l|=1\cr|k-l|\ge2\cr}\right.
\label{scalarp}
\end{equation} The first component of $\vec\alpha^g_k$  associated to the dilaton $\phi$ in
the original uncompactified theory Eq.(\ref{action})  is always zero. The corresponding
Dynkin diagram with $D-4$ nodes, which  from now on  we will refer to as  {\it the gravity line}
is depicted in Fig.1.
\vskip1cm
\hskip 2.5cm
\epsfbox{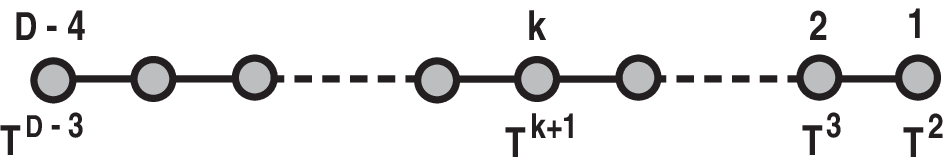}
\vskip.5cm

\begin{quote}
\begin{center}
\baselineskip 12pt {\small Fig.1.  The gravity line.

 Dynkin diagram of $A_{D-4}$ generated by the dimensional reduction to 3 dimensions.}
\end{center}
\end{quote} The remaining scalar corresponding to the missing simple root leading to  the
full
$A_{D-3}$ comes from dualising in three dimensions  the first vector (graviphoton) that
arises in the stepwise procedure namely the vector appearing already in $D-1$ dimensions.
The corresponding simple root is
\begin{equation}
\vec\alpha^{gp}= \sqrt{2} (0; (D-2) \beta_{D-1}, \beta_{D-2},
\beta_{D-3}, \dots ,\beta_3)\, .
\label{gravip}
\end{equation} Note  that this simple root $\vec\alpha^{gp}$ has a  non-vanishing scalar
product with $\vec\alpha^g_1$ (i.e with the simple root already appearing when
compactifying on $T^2$).  One has indeed 
$\vec\alpha^g_k \cdot \vec\alpha^{gp} = - \delta_{k,1}$.  Consequently it attaches itself to
the right of the gravity line.  The  other  ${1 \over 2} (D-4) (D-3)$ scalars coming from the
reduction of gravity down to three dimensions give  all the positive roots of $A_{D-3}$.

We now turn to theories with forms given by Eq.(\ref{action}) and begin by considering a
theory with only one $n_{A}$-form $F_{n_A}$ with dilaton coupling $a_A$ (and $n_A \leq
D/2$). Let us consider the first scalar arising  from the $n_A$-form upon
dimensional reduction up to $p=n_A-1$. The vector 
 $\vec\alpha_{n_A}^e$ associated to this scalar\footnote{The other scalars obtained by further  dimensional reduction  give
$\vec\alpha$-vectors that are linear combinations with positive integer coefficients of
$\vec\alpha_{n_A}^e$ and of the
 $\vec\alpha^g_k$.} will from now on be  called  {\it the would-be
electric root}. It is given by
\begin{equation}
\vec\alpha_{n_A}^e=   ({a_A \over \sqrt{2}};  \underbrace{b_{(n_A,D)}\, \beta_{D-1}, 
b_{(n_A,D)}\,
\beta_{D-2}, \dots ,b_{(n_A,D)}\, 
\beta_{D-n_A+1}}_{n_A-1 \, {\rm terms}},
\underbrace{0, \dots ,0}_{D-n_A-2 \, {\rm terms}})\, ,
\label{eroot}
\end{equation} with
\begin{equation} b_{(n_A,D)}= \sqrt{2} \, (D-n_A-1)\, .
\label{bdef}
\end{equation}First, we compute the scalar product of the would-be electric
root with the gravity line\footnote{ The scalar product of the  would-be electric root with
the graviphoton is one for any $D$ and $a_A$. This implies that when forms are present in a
theory with symmetry, the graviphoton is  never a simple root. Consequently we 
focus on the gravity line.}.  Using Eq.(\ref{gchain}) and Eq.(\ref{eroot}) we  find, {\it for any
D and dilaton coupling} $a_A$,
\begin{equation}
\vec\alpha^g_k\cdot\vec\alpha_{n_A}^e = -\delta_{k,n_A-1}\, .
\label{spg}
\end{equation} We then evaluate the length of the would-be electric root using 
Eq.(\ref{eroot}). We find
\begin{eqnarray} 
\vec\alpha_{n_A}^e \cdot\vec\alpha_{n_A}^e &=&  {(n_A-1)(D-n_A-1) \over (D-2)} + {a_A^2
\over 2}\, , \nonumber \\ &=& {(q^e_A+1)(q^m_A+1) \over (D-2)} + {a_A^2 \over 2}\, ,
\label{irredu}\\ &=& {\bar q}^{(e_A,m_A)} +1\, . \nonumber
\end{eqnarray} We thus see that the square length of the would-be electric root associated
to
$F_{n_A}$ can be written in terms of the intersection rule  equation Eq.(\ref{intrule}) giving
the intersection between the electric and the magnetic brane charged under $F_{n_A}$.

From now on we will restrict ourselves to simply laced groups. In  our normalisation  all their
roots are of square length two (see footnote 10).  In order for the would-be root to be a root,
one must  have
$\vec\alpha_{n_A}^e \cdot\vec\alpha_{n_A}^e=2$. Consequently   {\it the existence of a BPS
configuration in the original theory, consisting of an electric extremal
$p$-brane ($p \geq 1$) and its magnetic dual whose intersection is ${\bar
q}^{(e_A,m_A)}=1$,  is   a necessary
condition to have after dimensional reduction an enhanced simply laced Lie group symmetry.} 

Let us choose  the dilaton coupling to $F_{n_A}$  such that  ${\bar
q}^{(e_A,m_A)}=1$. Using the
scalar product  Eq.(\ref{spg}) of the would-be electric root with the gravity line,  we
can draw a would-be Dynkin diagram  where the
would-be electric root associated to  $F_{n_A}$  is connected to the $(n_A-1)^{th}$  node of the
gravity line as depicted in Fig.2.

In the dimensional reduction, there is an additional scalar  coming from the $F_{n_A}$ which
arises when one reaches $n_A+1$ non-compact dimensions.  It is obtained by
dualising the
$n_A$-form to a scalar. The corresponding   would-be
magnetic root  $\vec\alpha_{n_A}^m$ is given by
\begin{equation}
\vec\alpha_{n_A}^m=   (-{a_A \over \sqrt{2}};\underbrace{c_{n_A}\, \beta_{D-1},  c_{n_A}\,
\beta_{D-2}, \dots ,c_{n_A}\, 
\beta_{n_A+1}}_{D-n_A-1 \, {\rm terms}},
\underbrace{0, \dots ,0}_{n_A-2 \, {\rm terms}})\, ,
\label{mroot}
\end{equation} with
\begin{equation} c_{n_A}= \sqrt{2} \, (n_A-1)\, .
\label{cdef}
\end{equation} 
\vskip1cm
\hskip 3.5cm
\epsfbox{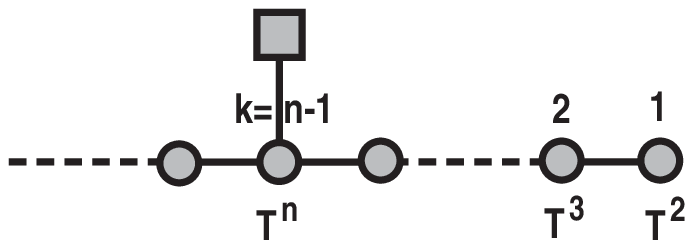}
\vskip .5cm
\begin{quote}\begin{center}
\baselineskip 12pt
 {\small Fig.2. Would-be Dynkin diagram.

 The would-be electric root associated to  $F_{n}$  is represented by a shaded square.}
\end{center} 
\end{quote} The scalar product of  the would-be magnetic root with the gravity line is
given by 
$\vec\alpha^g_k\cdot\vec\alpha_{n_A}^m = - \delta_{k,D-n_A-1}$ for any $D$ and
$a_A$. The square length of $\vec\alpha_{n_A}^m$ gives again the intersection rule
equations between the electric  and the magnetic brane namely
$\vec\alpha_{n_A}^m \cdot\vec\alpha_{n_A}^m =  {\bar q}^{(e_A,m_A)} +1$. Consequently
a necessary condition for  the would-be magnetic root to be a root is  the same as
previously.  This root may not be simple. We note that the scalar product of the would-be
magnetic root with the would-be electric root gives the intersection rule equation between
two {\it electric} branes
\begin{eqnarray} 
\vec\alpha_{n_A}^e \cdot\vec\alpha_{n_A}^m &=&  {(n_A-1)^2 \over (D-2)} - {a_A^2 \over
2}\, ,
\nonumber \\ &=& {(q^e_A+1)(q^e_A+1) \over (D-2)} - {a_A^2 \over 2}\, , \label{irm}\\ &=&
{\bar q}^{(e_A,e_A)} +1\, , \nonumber
\end{eqnarray}
and hence, from Eq.(\ref{qbar}), is integer if ${\bar q}^{(e_A,m_A)}=1$.
 
If we now generalise the previous discussion to a theory with more than one form $F_{n_A}$, 
we again find relations between the intersection rules and the would-be roots. For
instance considering a $F_{n_A}$ form and a $F_{n_B}$ form with $n_B > n_A$ the scalar product
of the two corresponding  would-be electric roots is given by
\begin{eqnarray} 
\vec\alpha_{n_A}^e \cdot\vec\alpha_{n_B}^e &=&  {(D-q^e_B-3)(q^e_A+1) \over (D-2)} +
{a_A a_B
\over 2}\, , \label{irab}\\ &=& {\bar q}^{(e_A,m_B)} +1\, , \nonumber
\end{eqnarray} which is the intersection rule between the magnetic extremal brane charged
under
$F_{n_B}$ and the electric extremal brane charged under $F_{n_A}$.

In this section we have  derived important relations between the existence of BPS
configurations of intersecting branes and the presence of a potential underlying symmetry
of a given  theory. In particular, we have shown that the existence of BPS
intersecting solutions   (with one common direction) between $e-m$ pairs of extremal
branes $(p \geq 1)$ for each form field strength  present in a theory given by
Eq.(\ref{action}) is a necessary condition to  have an enhanced symmetry corresponding to a
simply laced Lie algebra when the theory is dimensionally reduced.

We   briefly comment on the  non-simply laced groups. These algebras have long
and short roots.  If we normalise the long roots  to have square length two then the short
roots will have integer square length for all the non-simply  laced Lie algebra except
one\footnote{We thank Marc  Henneaux for a discussion on the specificity of $G_2$.} : $G_2$
where the short root has square length
$2/3$. In order for a theory given by Eq.(\ref{action}) to give a non-simply laced  symmetry
$B_n$,
$C_n$ or $F_4$ (except $G_2$) one must  have, as a necessary condition,   
intersecting BPS configurations with ${\bar q}=1$ or
$0$ for all the $e-m$ pairs of
branes of the theory.

\setcounter{equation}{0}

\section{Brane dynamics and symmetry}

In this section, using the result of the previous section and the opening of  branes
 described in Section 3, we will be able to reconstruct  all the oxidation end points described in
reference \cite{cjlp}  leading in three dimensions to all (split)
simply laced   group\footnote{The $A$-series corresponds to pure gravity
(without forms), which is the basic ingredient of our construction.} ${\cal G}$. 

Our starting point will be a theory given by Eq.(\ref{action}) in $D$  dimensions  with only
one
$n_A$-form field strength $F_{n_A}$ and its dilaton coupling $a_A$. We will  fix the dilaton
coupling such that there exists a zero-binding energy configuration between the
electric
$q^e_A$-brane ($q^e_A= n_A-2$)  and the magnetic $q^m_A$-brane with ${\bar
q}^{(e_A,m_A)}=1$. As explained in the previous section this is a necessary condition in
order to find  a new symmetry. Once the dilaton coupling of the 
form is fixed, we  require that, when the dimensionality of an intersection permits opening,
 the latter is consistent with charge conservation.  Namely we will impose that, if ${\bar
q}^{(e_A,m_A)}= q^e_A-1$,  the electric brane  open  on the magnetic brane.  As explained in
Section 3, this  requires the introduction of  a 
specific Chern-Simons term in  the action, which may contain a  new form field strength
$F_{n_B}$. The dilaton coupling $a_B$ of the new form
field strength is then again fixed modulo its sign by the necessary condition 
${\bar q}^{(e_B,m_B)}=1$.  The intersection rules between the branes corresponding to the
different forms can be calculated. We can then check if new openings are possible. If it is
the case, we  iterate the procedure until  consistency
of all the  openings are ensured. In this way, we will be able to  reconstruct all the
maximally oxidised theories corresponding to the simply laced groups. This leads  to the
following conclusion: {\it The existence of BPS configurations  with}
${\bar q}=1$ {\it between any electric extremal $p$-brane $(p \geq 1)$ and its magnetic dual,
along with the  requirement of consistency of brane opening in the
 original uncompactified theory  (characterised by at most one dilaton), is  a
necessary and sufficient condition to have a theory whose dimensional reduction 
down to three dimensions has a simple simply laced group ${\cal G}$ symmetry}.

We now turn to the proof.  We start
 with a theory in $D$ dimensions ($D \geq 3$) with one $n$-form field strength $F_n$ and
dilaton coupling $a_n$.  The dimension  of the electric brane corresponding to $F_n$ is
$q=n-2 \geq 1$ with $2q+4 \leq D$ (by convention the electric brane is always
smaller than the magnetic one). Requiring ${\bar q}^{(e,m)}=1$ and using Eq.(\ref{irredu})
we get, up to a sign, the dilaton coupling
\begin{equation} a_n^2(D) = 2 \frac{(1-q) D + (q^2+4q-1)}{D-2}\, .
\label{dilcon}
\end{equation}  We must have 
\begin{equation}
\label{condi} a_n^2 \geq 0\, .
\end{equation} 

For $q=1$, the condition Eq.(\ref{condi}) is satisfied for {\it any} dimension $D$.  The value of the
dilaton coupling is $a^2_3= 8/( D-2)$.  Thus we recover Eq.(\ref{dndilaton}). We get  the
model Eq.(\ref{dnaction}) characterised by an electric $1F$ and a magnetic $NS(D-5)$ with
intersections given by Eq.(\ref{dnbranes}). The BPS configuration between the
$e-m$ pair is given by a
$1F$ {\it living inside} the  $NS(D-5)$. No  opening can occur, henceforth there is no need
to add new form or Chern-Simons terms. The theory given by the action  Eq.(\ref{dnaction})
should then lead after dimensional reduction down to three to a coset model with $\cal G$
simply laced. This is  the case. Indeed, the would-be electric root and the would-be 
magnetic root associated to $H_3$ are simple roots  and we find  the $D_n$-series with
$D=n+2$. 

For $q >1$ Eq.(\ref{condi}) leads to the following constraint on $D$,
\begin{equation} 2q+4 \leq D \leq q+5 +\frac{4}{q-1}\, .
\label{condid}
\end{equation} The constraint Eq.(\ref{condid}) can only be satisfied for $q=2$ and $q=3$. 
We have  5 possibilities 
\begin{eqnarray} q=2 \qquad &{\rm and}& \left\{ \begin{array}{l} D=11 \\ D=10 \\D=9 \\D=8 
\end{array} \right.\nonumber \\ q=3 \qquad &{\rm and}& D=10.
\label{condiq}
\end{eqnarray} For  $q=2$ ($F_4$) the electric brane can always
potentially open on the magnetic dual.

\underline{$q=2$, $D=11$}

In this case Eq.(\ref{dilcon}) gives $a_4=0$, namely  no dilaton coupling. There is a 2-brane
and its dual magnetic  5-brane  with  the   intersection $5 \cap 2=1$. To open the
2-brane  on the 5-brane,  it is necessary to add a Chern-Simons term. Here we
are in the special case
$D=3q+5$ and there is no need to add a new field strength to build the Chern-Simons term.
This term is proportional to $ F_4 \wedge F_4 \wedge A_3$.  Our construction procedure
stops here  as   no new form are introduced.  This theory is expected, from Section 4,  to
have a symmetry $E_8$ after dimensional reduction to three dimensions.  This is the
case. The intersection of the 2-brane with  the 5-brane along one
common dimension, together with the consistency condition on openings, has fixed  the action 
to be that of eleven dimensional supergravity whose  enhanced symmetry in the
dimensional reduction  down to three is  indeed $E_8$ \cite
{marcuschwarz}. 

\underline{$q=2$, $D=10$}

In this case, Eq.(\ref{dilcon}) gives\footnote{ The plus sign here  is just a
matter of convention only the relative signs between dilaton couplings matters.}
$a_4=1/2$. There is a 2-brane that we denote, anticipating the result, 
$D2$, and its magnetic dual which is a 4-brane, $D4$.  
According to our logic, we demand that the opening of the $D2$ on the $D4$  be consistent. 
This
requires the introduction of a Chern-Simons term
proportional to
$ A_3\wedge F_4 \wedge H_3$ where  $H_3$ is a new field strength, giving a new electric 
1-brane  (that we call $1F$),  with its  associated dilaton coupling $a_3$. 
From  Eq.(\ref{dilcon}) one finds
$a_3=-1$. The minus sign is imposed by the requirement that  the intersection rules
between the $1F$ and the $D4$ gives a BPS configuration, namely $1F \cap D4 =0$. 
The $1F$ can potentially open on the $D2$ in virtue of the  intersection rule $1F \cap D2 =0$. 
Consistency of this opening implies the introduction of a field strength $F_2$ through a 
Chern-Simons type  term in
the action leading to 
$ d*H_3 \propto * F_4 \wedge F_2$. Note that this is an opening of an electric brane on another
electric brane.
The branes corresponding to this new
$F_2$ are a 0-brane ($D0$) and a 6-brane ($D6$). The corresponding dilaton coupling
$a_2^2$ is again fixed by Eq.(\ref{dilcon}) and the sign is fixed by requiring that the intersection
rule for $1F$ and $D0$ yields a BPS configuration, namely  $1F \cap D0 =0$. One gets
$a_2=3/2$. Having fixed all the dilaton couplings, it is then possible to check all the
 intersection rules and to
introduce a  Chern-Simons like term for each possible opening.  In this way one checks that no
other forms are required and   one recovers type $IIA$ theory.

In our constructive approach, one thus recovers   a theory
having    after  reduction down to three dimensions a symmetry which is again, as expected,
$E_8$. Being the dimensional reduction of 11 dimensional supergravity, it is not an oxidation
endpoint.

\underline{$q=2$, $D=9$}

Eq.(\ref{dilcon}) now gives $a_4=2 / \sqrt{7}$.  There is an electric
2-brane (noted ${\bf 2}$) and its dual magnetic 3-brane  (noted ${\bf 3}$) with the following
pairwise BPS configurations read off, from Eq.(\ref{intrule}),
\begin{equation} {\bf 2} \cap {\bf 2}=0 \qquad {\bf 3} \cap {\bf 3}=1
\qquad {\bf 2} \cap {\bf 3}=1\, .
\label{e7ir1}
\end{equation} The opening of  ${\bf 2}$ onto the `host'
{\bf 3} requires the introduction  a new form $F_2$ with $a_2^2=16 / 7$ and a Chern-Simons
term in the action  proportional to 
\begin{equation} F_4 \wedge A_3 \wedge F_2 \qquad {\rm with} \qquad F_4 =dA_3\, .
\label{cse7}
\end{equation}  One gets a new electric 0-brane (noted ${\bf 0}$) and a new magnetic 
5-brane (noted ${\bf 5}$). There is one  BPS-configuration possible associated to this
form, namely
${\bf 5} \cap {\bf 5}=3$. The sign of $a_2$ is chosen to allow  new  BPS configurations
between  branes charged under different forms. Picking $a_2 =-4 / \sqrt{7}$ one gets the
following additional BPS intersections,
\begin{eqnarray} {\bf 2} \cap {\bf 0}=0 \qquad &{\bf 3} \cap {\bf 0}=-1&
\qquad {\bf 5} \cap {\bf 3}=3\, ,
\label{e7ir2}\\ &{\bf 2} \cap {\bf 5}=1\, .& 
\label{e7ir3}
\end{eqnarray} The last intersection Eq.(\ref{e7ir3})  yields the possible opening of ${\bf 2}
\rightarrow {\bf 5}$  whose consistency is precisely ensured by the Chern-Simons term 
already introduced in  Eq.(\ref{cse7}).  There is no more possible opening hence our
constructive procedure stops here. This theory is expected, from Section 4, to have a
symmetry $E_7$ after dimensional reduction to three dimensions.  This is indeed the case! It
corresponds to the  theory which is the oxidation  endpoint of ${\cal G} =E_7$  
\cite{cjlp}.

\underline{$q=2$, $D=8$}

The dilaton coupling $a_4$ is, from   Eq.(\ref{dilcon}),
$a_4= -1$ (the minus sign is purely conventional). There is an electric 2-brane  ${\bf
2_e}$ and its magnetic dual which is also a 2-brane,  ${\bf 2_m}$. One has the following
BPS-configurations,
\begin{equation} {\bf 2_e} \cap {\bf 2_e}=0 \qquad {\bf 2_m} \cap {\bf 2_m}=0
\qquad {\bf 2_e} \cap {\bf 2_m}=1\, .
\label{e6ir1}
\end{equation} The opening of ${\bf 2_e} \rightarrow {\bf 2_m}$ requires the introduction of
a Chern-Simons terms in the action of the following form,
\begin{equation}  A_3 \wedge F_4 \wedge F_1 \qquad {\rm with} \qquad F_4 =dA_3\, ,
\label{cse6}
\end{equation}  hence the introduction of a new one-form $F_1$ and the corresponding
$(-1)$-electric brane, $\bf{ -1}$, and its dual magnetic 5-brane, $\bf {5}$. The
dilaton coupling is, from Eq.(\ref{dilcon}), 
$a_1^2=4$. Its sign  is fixed to be plus by the requirement that there exists a further
BPS-configuration whose opening is also consistent with  the Chern-Simons term
Eq.(\ref{cse6}), namely,
\begin{equation} {\bf 2_e} \cap {\bf 5}=1\, .
\label{e6ir2}
\end{equation}With the introduction of $F_1$, and $a_1$ fixed to be $2$, there are no
further possible openings and our building procedure stop here.  This theory is expected,  from
Section 4, to have a symmetry $E_6$ after dimensional reduction to three dimensions.  This is
indeed the case!  It corresponds to the  theory which is the oxidation  endpoint of ${\cal G}
=E_6$.

\underline{$q=3$, $D=10$}

One has a five form field strength $F_5$ and  from Eq.(\ref{dilcon}), $a_5=0$,  namely
no dilaton coupling. In order to have a symmetry and avoid a degeneracy  in the
brane spectrum we impose a self-duality condition  on $F_5$.  One has a 3-brane (${\bf 3}$)
and the following BPS configuration,
\begin{equation} {\bf 3} \cap {\bf 3}=1\, .
\label{e7ire}
\end{equation} There is no possible opening, thus no need to  add new forms. This theory in
$D=10$ dimensions with a self-dual 5-form   gives indeed, again after dimensional reduction
down to three, 
${\cal G}=E_7$ \cite{cjlp}. It constitutes an oxidation   endpoint if we allow for the self-duality
condition.

\setcounter{equation}{0}

\section{Intersecting branes in ${\cal G}^{+++}$}

In this section we analyse the moduli space of the intersecting brane solutions of the
theories discussed above, which  are the oxidation endpoints for
the simple simply laced Lie groups $\cal G$. 

Recall that these solutions were obtained by enforcing the ansatz Eq.(\ref{ansatz}) whose
significance we now discuss. We rewrite this equation in terms of the moduli
\begin{eqnarray}
\label{moduli}
  p^1& =& \ln B\nonumber\\
 p^a &=& \ln G\qquad\qquad a=2,\dots, D-p\\ p^{D-p+i}&=&\ln C_i\qquad\qquad i=1,\dots,
p\nonumber
\end{eqnarray} and we get
\begin{equation}
\label{branemetric} (p+3-D)\,   p^a=  p^1+ \sum_{i=1}^p   p^{D-p+ i}\qquad a=2,\dots, D-p\, .
\end{equation} For the extremal intersecting brane solutions Eqs.(\ref{metric_sol}) and
(\ref{dilaton_sol}) one has
\begin{eqnarray}
 p^1 &=& - \sum_A \frac{D-q_A-3}{\Delta_A}  \ln H_A\, , \nonumber\\
 p^a &= &\sum_A \frac{q_A+1}{\Delta_A} \ln H_A\, , \nonumber\\ p^{D-p+i}&=&- \sum_A
\frac{\delta^{(i)}_A}{\Delta_A}\ln H_A\, , \nonumber\\
\phi &=& \sum_A \frac{D-2}{\Delta_A}\varepsilon_A a_A \ln H_A\, . \label{phicond}
\end{eqnarray} For simply laced groups, ${\Delta_A}/(D-2) =2$. Taking into account the
intersection rule equation Eq.(\ref{intrule}) and the ansatz Eq.(\ref{branemetric}), one gets
\begin{equation}
\label{gplus}
\sum_{\alpha=1}^D (  p^\alpha)^2 -\frac{1}{2}(\sum_{\alpha=1}^D   p^\alpha)^2 +
\frac{1}{2}
\phi^2=\frac{1}{2}\sum_A \ln^2 H_A(x^a)\, .
\end{equation} The relations Eqs.(\ref{branemetric}) and (\ref{gplus}) have a
group-theoretical significance which we shall uncover. This section is based on
reference \cite{ehtw} to which the reader is referred for more detailed discussions.

\hskip 2.5cm
\epsfbox{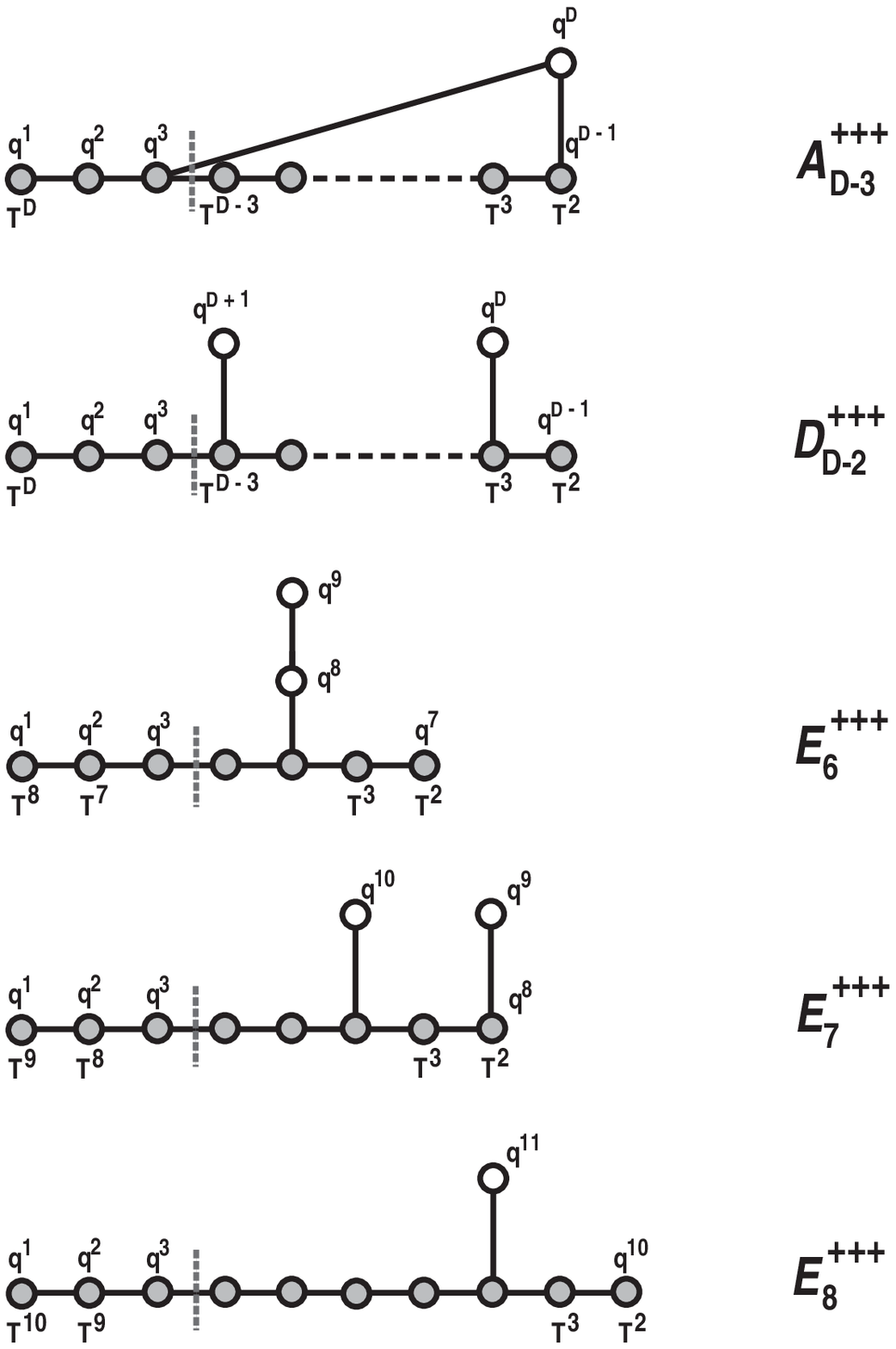}
\vskip .5cm
\begin{quote}\baselineskip 12pt \begin{center}
 {\small Fig.3.  Dynkin diagram of simply laced Kac-Moody algebras ${\cal G}^{+++}$.}
\end{center} {\small  The nodes of the gravity line are shaded. The nodes characterised by
the Chevalley parameters
$q^1,q^2,q^3$ are the Kac-Moody extensions of the Lie algebras $\cal G$ whose Dynkin
diagrams are depicted by the nodes on the right of the vertical dashed lines. The triple
extension $A^{+++}_{D-3}$, stemming from pure gravity, is included for sake of generality.}
\end{quote}

We first recall that the Lie algebra
$\cal G$ can be embedded in a very extended Kac-Moody algebra ${\cal G}^{+++}$. The
simple roots of ${\cal G}^{+++}$ are given by adding  two nodes to the gravity line of  the
Dynkin diagram of the affine extension $\cal G^+$ of $\cal G$, thus increasing by three the
rank of
 $\cal G$.  The resulting Dynkin diagrams for simply laced algebras are shown in Fig.3. 

The group
$SL(D)$ defined by this triple extended gravity line can be extended to the  full
deformation group $GL(D)$ whose algebra, generated by  $D^2$ generators
$K^\alpha_{~\beta}~,~\alpha, \beta=1,\dots, D$ , is a subalgebra of  
${\cal G}^{+++}$. The $K^\alpha_{~\beta}$ satisfy the following commutation relations
\begin{equation}
\label{Kcom} [K^\alpha_{~\beta},K^\gamma_{~\delta}]
=\delta^\gamma_{~\beta}K^\alpha_{~\delta}-\delta^\alpha_{~\delta}K^\gamma_{~\beta}\,  .
\end{equation} One considers the Cartan subalgebra of ${\cal G}^{+++}$ generated by the
$K^\alpha_{~\alpha}$ and $q= r-D$ abelian generators $R_u$ where $r$ is the rank of ${\cal
G}^{+++}$. We write the corresponding abelian group element $g$ as
\begin{equation}
\label{explicit} g=\exp (\sum_{\alpha=1}^D p^\alpha K^\alpha{}_\alpha) \exp (\sum_{u=1}^q
\phi^u R_u)\, .
\end{equation} The group parameters $\{p^\alpha, \phi^u\} $ are related, for diagonal metric
field configurations, to  the metric fields $g_{\alpha\alpha}$  by
\begin{equation}
\label{metricb} g_{\alpha\alpha}= e^{2p^\alpha}\eta_{\alpha\alpha}\, ,
\end{equation} where $\eta_{\alpha\alpha}$ is the Minkowskian metric $(-,+,+,\dots,+) $. The 
$\phi^u$ are identified to dilaton fields.  For the simply laced groups considered here there
is at most one dilaton.
 It was  shown in reference \cite{ehtw} that compactification of this theory on a k-torus in a flat
background could be interpreted as the embedding of the algebra
${\cal G}^{(k)}$ into ${\cal G}^{+++}$ obtained by deleting, starting from the left, 
$D-k$ nodes of its gravity line together with  nodes attached to the deleted ones.
Rewriting the group element Eq.(\ref{explicit})  in the Chevalley basis of 
 ${\cal G}^{+++}$ as $g=e^{q^mH_m}$, it amounts to set $q^m=0$ for $m=1, \dots, D-k$ along
with
 $q^x$ where $x$ labels a possible node attached to one of the $q^m$ (see Fig.3). 
This embedding is characterised by Eq.(3.29) of reference
\cite{ehtw} which can be written as
\begin{equation}
\label{compact} ( k+2-D)\, p^a=  \sum_{i=1}^k p^{D-k+i}\, ,\qquad a=1,\dots, D-k
\end{equation} where the $p^\alpha=\{p^a,p^{D-k+i}\}$ are related to the background metric
through Eq.(\ref{metricb}). For such an embedding the group $S({\cal G}^{(k)})$ generated by
the Weyl generators  of ${\cal G}^{(k)}$ and by the outer automorphisms of its Dynkin
diagram is a subgroup of the group 
$S({\cal G}^{+++}) $ which leaves invariant the quadratic form
\begin{equation}
\label{gp}
\sum_{\alpha=1}^D (p^\alpha)^2 -\frac{1}{2}(\sum_{\alpha=1}^D p^\alpha)^2 + \frac{1}{2}
\phi^2\, .
\end{equation} Taking $k=p+1$ in Eq.(\ref{compact}) and interchanging the time modulus
$p^1$ with $p^{D-p}$ one recovers the ansatz Eq.(\ref{branemetric}).  The interchange of
moduli is a Weyl transformation in
${\cal G}^{+++}$. Thus the ansatz  Eq.(\ref{branemetric}) defines an embedding ${\cal
G}^{(p+1)}$ in ${\cal G}^{+++}$ conjugate by Weyl reflection to the one defined by 
Eqs.(\ref{compact}) and (\ref{gp}), provided the right hand side of Eq.(\ref{gplus}) is
constant under the  group $S({\cal G}^{(p+1)})$. This  is ensured if time is compactified
along with the original  $p$ dimensions. When there exists  an underlying string theory
associated to the ${\cal G}^{+++}$ theory  and  an  interpretation of Weyl transformations in
terms of T-dualities \cite{ehtw}, it amounts to allow T-duality along the time direction
\cite{tdual}.  One can show, using the formalism of reference
\cite{ehtw} that the embedding Eq.(\ref{branemetric}) expressed in the Chevalley basis 
corresponds to the conditions
$q^1=q^2=\dots=q^{D-p}$. 

It is of interest to compare for a given maximally oxidised theory the embedding of ${\cal
G}^{(p+1)}$ in ${\cal G}^{+++}$  describing the intersecting brane solutions with the
embedding of the overextended Kac-Moody algebra
${\cal G}^{++}$ in ${\cal G}^{+++}$ that characterises the cosmological Kasner solutions.
These can be written as
\begin{eqnarray}
\label{ukasner} ds^2= -e^{2 p^1  (t)}dt^2 +
\sum_{\alpha=2}^D e^{2 p^\alpha (t)}~(dx^\alpha)^2 &; &p^\alpha (t)=\tilde p^\alpha t
\quad\hbox{with}\quad
\tilde p^\alpha=\hbox {constant}\, ,\quad 
\\
 \label{udilat} p^1&=&\sum_{\alpha=2}^D      p^\alpha	\, ,
\end{eqnarray} together with Eq.(\ref{gplus}), with the right hand side put equal to zero
\cite{ehtw}.
  The embedding of this overextended Kac-Moody algebra ${\cal G}^{++}$  is defined by
$q^1=0$ or equivalently by Eq.(\ref{udilat}). Its physical relevance was put into evidence by
its occurrence in the cosmological billiards describing the evolution of a
universe in the vicinity of a cosmological singularity \cite{bill1} for all maximally oxidised
theories
\cite {bill2}. The embedding of ${\cal G}^{(p+1)}$ in ${\cal G}^{+++}$ is also an embedding in
its subalgebra ${\cal G}^{++}$ defined by $q^1=q^2$ which is conjugate to the one defined by
$q^1=0$. Hence, although both the algebra spanned by intersecting branes solutions and by
the cosmological billiard solutions  are both subalgebra of the very extended Kac-Moody
algebra
${\cal G}^{+++}$, the first one cannot be embedded in the second and the common origin of
their symmetry is only revealed at the level of the very extended Kac-Moody algebra.

It is also possible to interpret the embedding Eq.(\ref{compact}) in terms of extremal
intersecting branes where time is transverse to all branes. Such intersecting brane
solutions
\cite{ah} indeed exist  in  exotic theories \cite{exop}. Their role in the context of full ${\cal
G}^{+++}$ invariance will not be discussed here.

\section*{Acknowledgments}

We are very grateful to Anne Taormina for collaboration at
early stages of this work.  Laurent Houart is greatly  indebted to
Riccardo Argurio for discussions on intersection rules and BPS states beyond supersymmetry.
He would like also to thank the CECS  (Centro de Estudios) of  Valdivia (Chile)  and the Pontificia
Universidad Catolica de Chile for the warm  hospitality extended to him while part of this work
was done. Peter West thanks the Universit\'e Libre de Bruxelles for the
 stimulating atmosphere during his visits.   

This work was supported in part  by the NATO grant PST.CLG.979008,
  by the ``Actions de Recherche Concert\'ees'' of the ``Direction de la Recherche
Scientifique - Communaut\'e Fran\c caise de Belgique, by a ``P\^ole d'Attraction
Interuniversitaire'' (Belgium), by IISN-Belgium (convention 4.4505.86), by Proyectos
FONDECYT 1020629, 1020832 and 7020832 (Chile) and by the European Commission RTN
programme HPRN-CT00131, in which F.~E. and L.~H. are associated to the Katholieke
Universiteit te Leuven (Belgium).

\end{document}